\input{aipcheck}

\documentclass[
    ,final            
  ]
  {aipproc}

\layoutstyle{6x9}
\newcommand{\met} {\mbox{$\not\!\!E_T$}}

\begin{document}

\title{Search for trilepton SUSY signal at CDF}


\author{John Strologas}{
  address={New Mexico Center for Particle Physics, University of New Mexico, Albuquerque, NM 87131, USA\\
(for the CDF collaboration)}
}

\begin{abstract}
The chargino-neutralino production with subsequent leptonic decays
is one of the most promising supersymmetry (SUSY) signatures at the Tevatron 
proton-antiproton collider. We present the most recent results on the search 
for the three-lepton and missing-transverse-energy SUSY signature using 3.2 fb$^{-1}$
of data collected with the CDF II detector. The results are interpreted within the minimal 
supergravity (mSUGRA) scenario. 
\end{abstract}

\maketitle


\section{Introduction}

Theories with supersymmetry (SUSY) are arguably the most compelling 
of those that attempt to resolve the outstanding particle-physics questions
\cite{BaerTata}.  Starting from the simple 
hypothesis of the existence of an identical fermion for every known boson 
and vice-versa, we can resolve remaining asymmetry problems (fermion/boson,
particle/antiparticle), allow for the 
exact unification of interactions, and solve the fine-tuning problem.  If 
$R$-parity is conserved, an axiom that is often invoked to avoid proton decay, 
then the lightest supersymmetric particle (LSP) will 
be stable, providing an excellent candidate for cold dark matter.  
Moreover, SUSY offers a radiative electroweak-symmetry breaking capability.

Superpartners of the known standard-model (SM) particles, differing only in spin, have 
not been observed and have been ruled out.  As a result, SUSY should be a broken 
symmetry.  Our working assumption is a minimal supersymmetric standard model
(MSSM) with soft SUSY breaking.  If the transfer of SUSY-breaking to the visible 
sector happens through gravity, the SUSY scenario is SUGRA and the LSP is the 
lightest neutralino.  If the transfer takes place through gauge fields, the SUSY 
scenario is the GMSB and the LSP is the gravitino. 
In mSUGRA, the only free parameters of the theory are
the common scalar mass $m_0$, the common gaugino mass $m_{1/2}$, the ratio
of Higgs vacuum expectation values tan$\beta$, the trilinear sfermion-sfermion-Higgs
coupling $A_0$ and the the sign of the higgsino scale parameter $\mu$.
In this paper we interpret our result in the mSUGRA scenario, although 
the search is performed in a scenario-independent manner.  

\section{Chargino-Neutralino production and decay}

The fermionic partners of the SM gauge bosons and the Higgs, the gauginos and
the higgsinos, mix to give two observable charginos and four observable neutralinos.
Assuming $R$-parity conservation, the charginos and neutralinos have to be produced
in pairs.  Due to the nature of the gaugino and higgsino mixing under the assumption of gaugino mass
unification, the highest cross section at the Tevatron is that of the 
associated production of the lightest chargino and the 
next-to-lightest neutralino.  The production takes place mainly 
through an off-shell $W$ boson, since the $t$-channel production with a squark 
propagator is unfavored due the high squark mass limits.

The charginos and neutralinos decay either through sleptons, which always
decay to leptons, or through off-shell gauge bosons, which decay to 
leptons only a fraction of the time.  In any case, the leptonic decays of
the chargino-neutralino pair will result in three leptons and missing transverse
energy ($\met$) from the undetected neutrinos and lightest neutralino (LSP).
We investigate the leptonic decays because the SM trilepton backgrounds are 
very low.  Given that the chargino-neutralino production cross sections of the order of 
0.1-1 pb (depending on the SUSY parameter space) have not been excluded yet, and given the 
current mass limits for squarks and gluinos ($> 300$ GeV/$c^2$), the trilepton signature is 
the ``golden'' channel for the discovery of supersymmetry at the Tevatron \cite{baer}.

In order to estimate the SUSY signal for a particular mSUGRA point, we determine the 
mSUGRA mass spectrum and decay branching ratios using {\sc isasugra} \cite{isasugra} 
and we simulate the events using Monte Carlo ({\sc Pythia} generator \cite{pythia} 
and CDF detector simulation).  We normalize the events using production cross sections 
determined with {\sc prospino} \cite{prospino2}.

\section{Standard-model trilepton backgrounds}

The main sources of dileptons at the Tevatron are the Drell-Yan (DY) process
and the semileptonic heavy-flavor quark ($b$, $c$) decays, the latter being 
significant at low dilepton invariant masses ($M_{\ell\ell}<35$ GeV) \cite{strologas}.
The main sources of trileptons are the diboson leptonic decays and
the above dilepton production with the addition of a photon (that converts) 
or a fake lepton, i.e., a jet that fakes an electron or a track that fakes a muon.  
The source of fakes is hadrons (mainly kaons and pions) that decay-in-flight or
punch-through; we thus associate the fakes to light-flavor quarks.  The $\met$ comes from 
neutrinos or from limited energy resolution or from event mis-reconstruction.  At higher $M_{\ell\ell}$ and $\met$, trileptonic $t\bar{t}$ signal is also considered. 

The electroweak backgrounds (diboson, DY+$\gamma$) as well as $t\bar{t}$ are estimated
using Monte Carlo simulation.  In this analysis we eliminate the heavy-flavor 
QCD by applying appropriate dilepton-mass, lepton-$p_T$ and $\met$ cuts.  The light-flavor QCD (fakes) 
is estimated from CDF data, by selecting two leptons and applying a fake-rate on 
additional jets or tracks present in the event.  These fake-rates are measured as a function
of the fakeable jet $E_T$ or track $p_T$ using jet-rich CDF data.

\section{Analysis strategy and event selection}

The most critical part of any new-physics search is the accurate estimation
of the SM backgrounds. For this purpose we define dilepton and trilepton
control regions -- kinematically orthogonal to our signal region -- 
where we confirm good understanding of the backgrounds.  We look at the signal region 
only after confirming that the CDF data and SM expectation agree in the control regions 
in both event yields and kinematic distributions, thus performing a statistically unbiased 
analysis.  

We analyze 3.2 fb$^{-1}$ of CDF data, collected up to the summer of 2008.
We utilize the low-$p_T$ ($> 4$ GeV/$c$) dielectron/dimuon and the
high-$p_T$ ($> 18$ GeV/$c$) single electron and single muon triggers.
Our leptonic objects are high-quality isolated central electrons and muons, 
whereas the third leptonic object can be an isolated central track.  
The leptons are isolated if the extra energy in a cone of $\Delta R = 0.4 $ around them
is less than 10\% of their energy. 
To maximize sensitivity, 
we investigate tight and loose lepton channels separately.  This analysis \cite{rob}
is the update of a previous CDF search \cite{sourabh}.

The signal region is defined as trileptons with $M_{\ell\ell}>20$ GeV/$c^2$ (for
reduction of photonic DY and heavy-flavor QCD), $M_{\ell\ell}<76$ GeV/$c^2$ or $M_{\ell\ell}>106$ GeV/$c^2$
(for reduction of $Z$ boson resonances), $\met>20$ GeV (for DY/QCD reduction and
increase of signal sensitivity), and low jet activity (for $t\bar{t}$ and QCD reduction).
The $p_T$ of the leading lepton is $>15$ or $>20$ GeV/$c$ (depending on the channel)
whereas the two subleading leptons 
can be as low as 5 GeV/$c$.
After this selection, the remaining trilepton SM backgrounds in the signal region
are diboson (61\%), DY+$\gamma$ (22\%) and QCD (15\%).

The control regions are defined in the $M_{\ell\ell}$ vs. $\met$ phase space,
for both dilepton and trileptons (for the latter case, at least one of the signal kinematic 
cuts has to be inverted).  
The $Z$-boson dilepton control regions validate the DY background,
the trigger and lepton-identification efficiencies, and our knowledge of the luminosity.  
The low-mass low-$\met$ control regions validate the QCD backgrounds.  The high jet multiplicity 
regions validate the $t\bar{t}$ background.

\section{Results and current work}

Figure 1 shows the comparison between background expectation and CDF dilepton and trilepton
data in our control regions.  
\begin{figure}[!]
\begin{minipage}[!]{7cm}  
\begin{center}
\includegraphics[scale=.38]{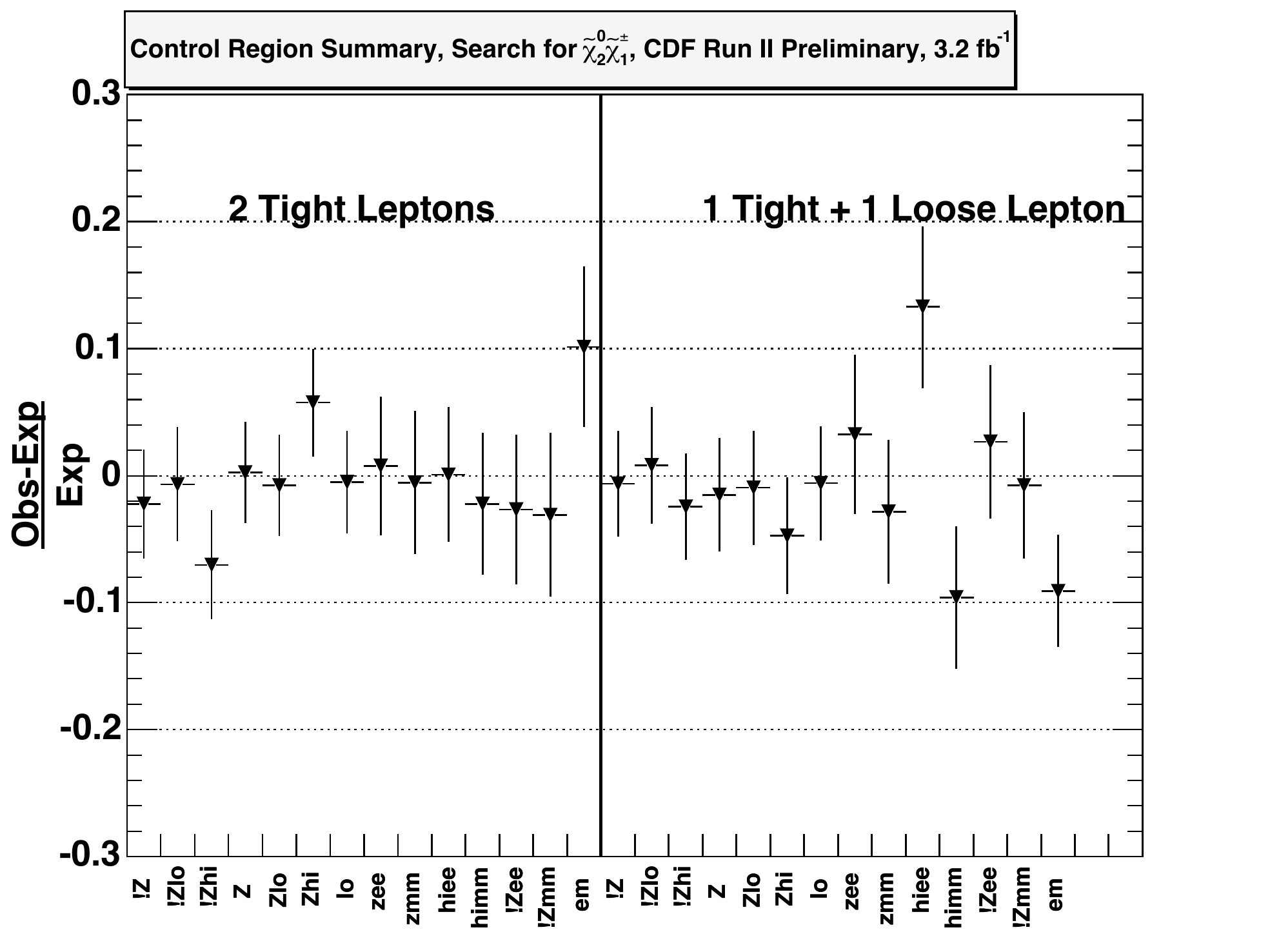}
\end{center}
\end{minipage}  
\hfill
\begin{minipage}[!]{7cm} 
\begin{center}
\includegraphics[scale=.395]{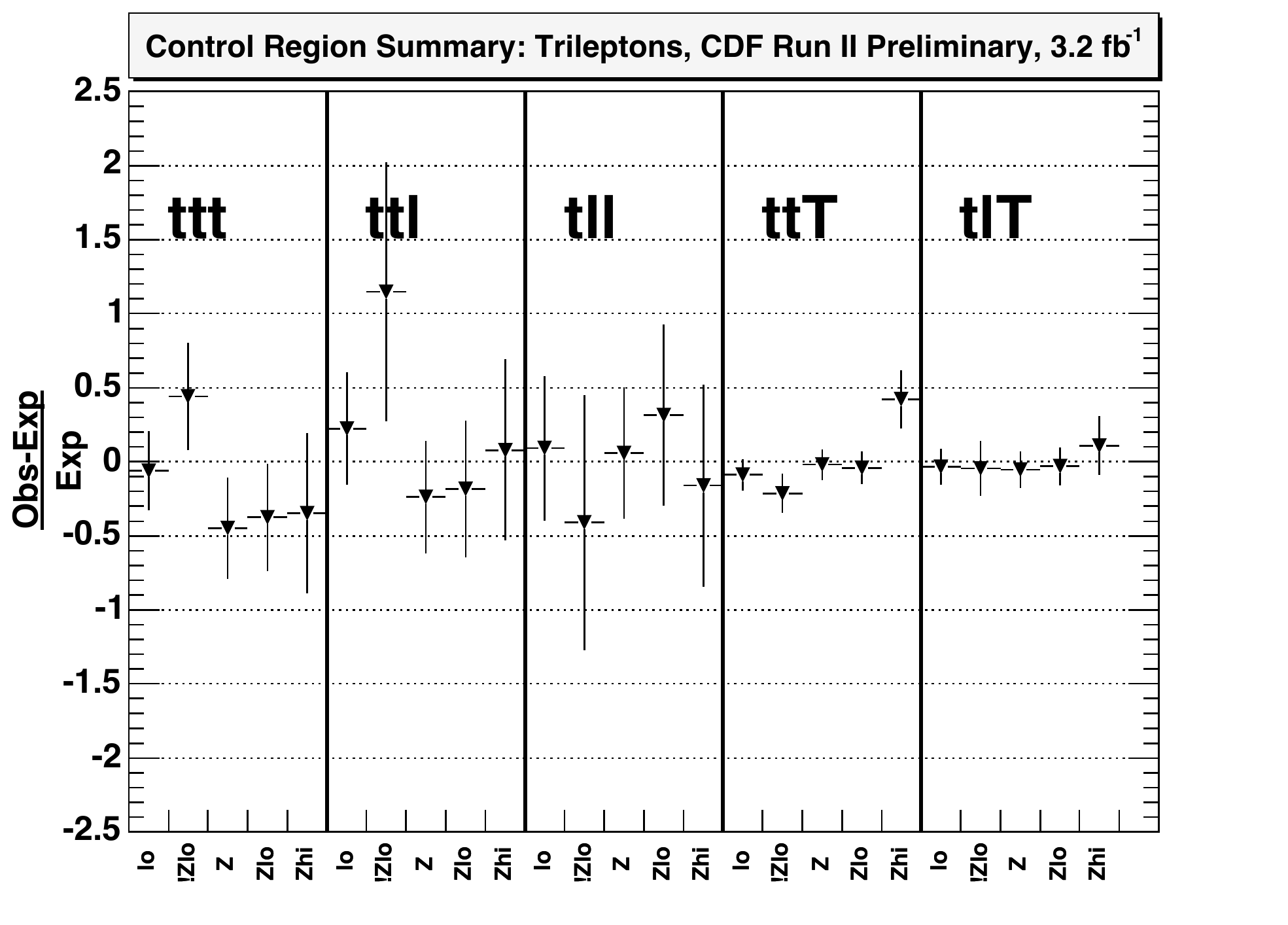}
\caption{Comparison between observation and expectation in the dilepton (left) and trilepton (right) control regions, for tight (t) and loose (l) leptons and tracks (T).}
\end{center}
\end{minipage}  
\hfill
\end{figure}
After observing excellent agreement in the control regions,
we look at the signal region.  There, we expect 
$1.5 \pm 0.2$ SM trilepton and $9.4 \pm 1.4$ SM dilepton+track
events and we observe 1 and 6 events respectively.
Our results are consistent with SM predictions,
so we proceed to setting limits in the mSUGRA scenario.

First, we set $m_0=60$ GeV/$c^2$, $\tan\beta=3$ and $\mu>0$ and vary $m_{1/2}$ to investigate a range
of chargino masses from 98 to 174 GeV/$c^2$.  The expected and observed excluded limits in cross 
section times branching ratio vs. chargino mass can be seen in Figure 2.
\begin{figure}[!]
\begin{minipage}[!]{7cm}  
\begin{center}
\includegraphics[scale=.38]{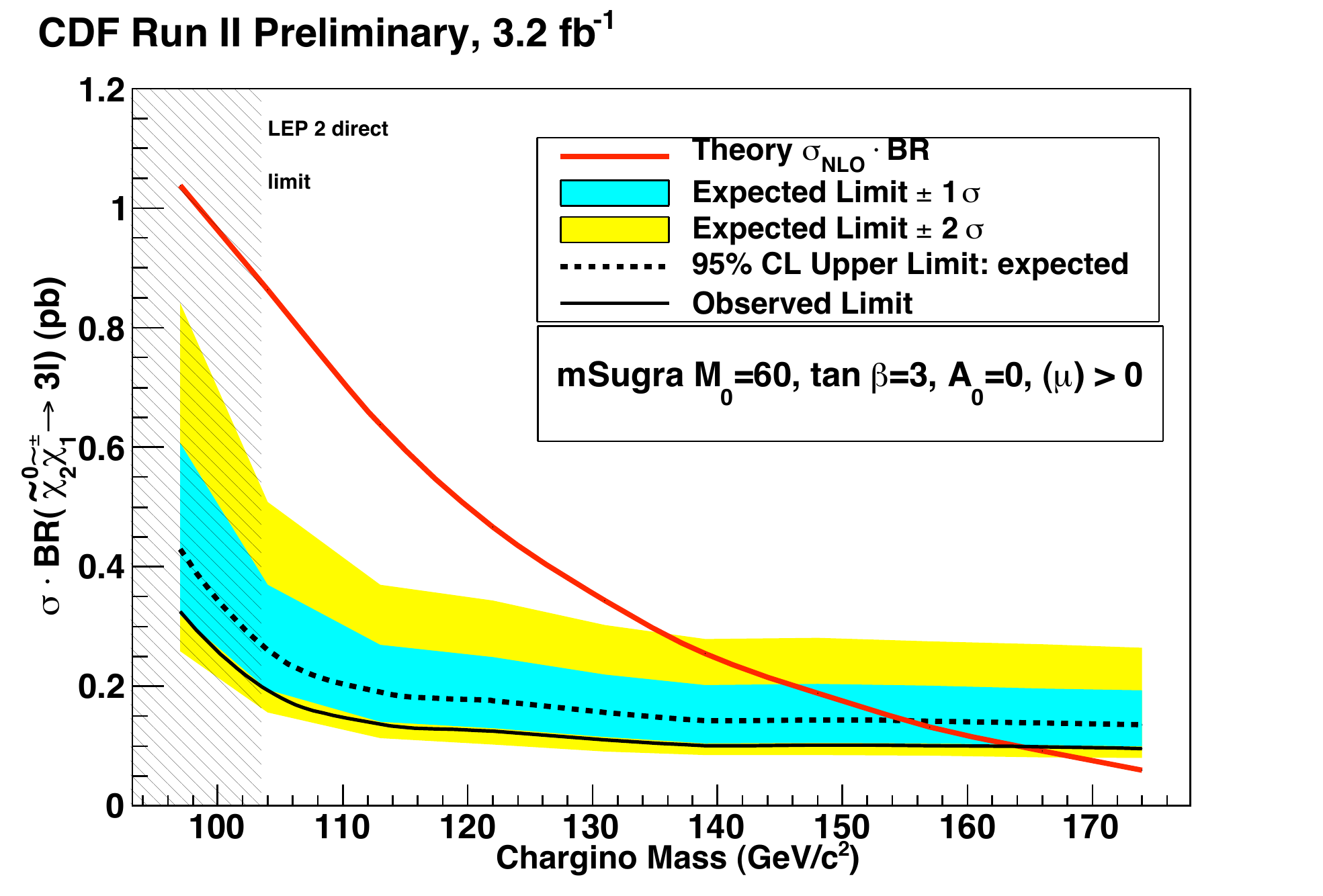}
\end{center}
\end{minipage}  
\hfill
\begin{minipage}[!]{7cm} 
\begin{center}
\includegraphics[scale=.4]{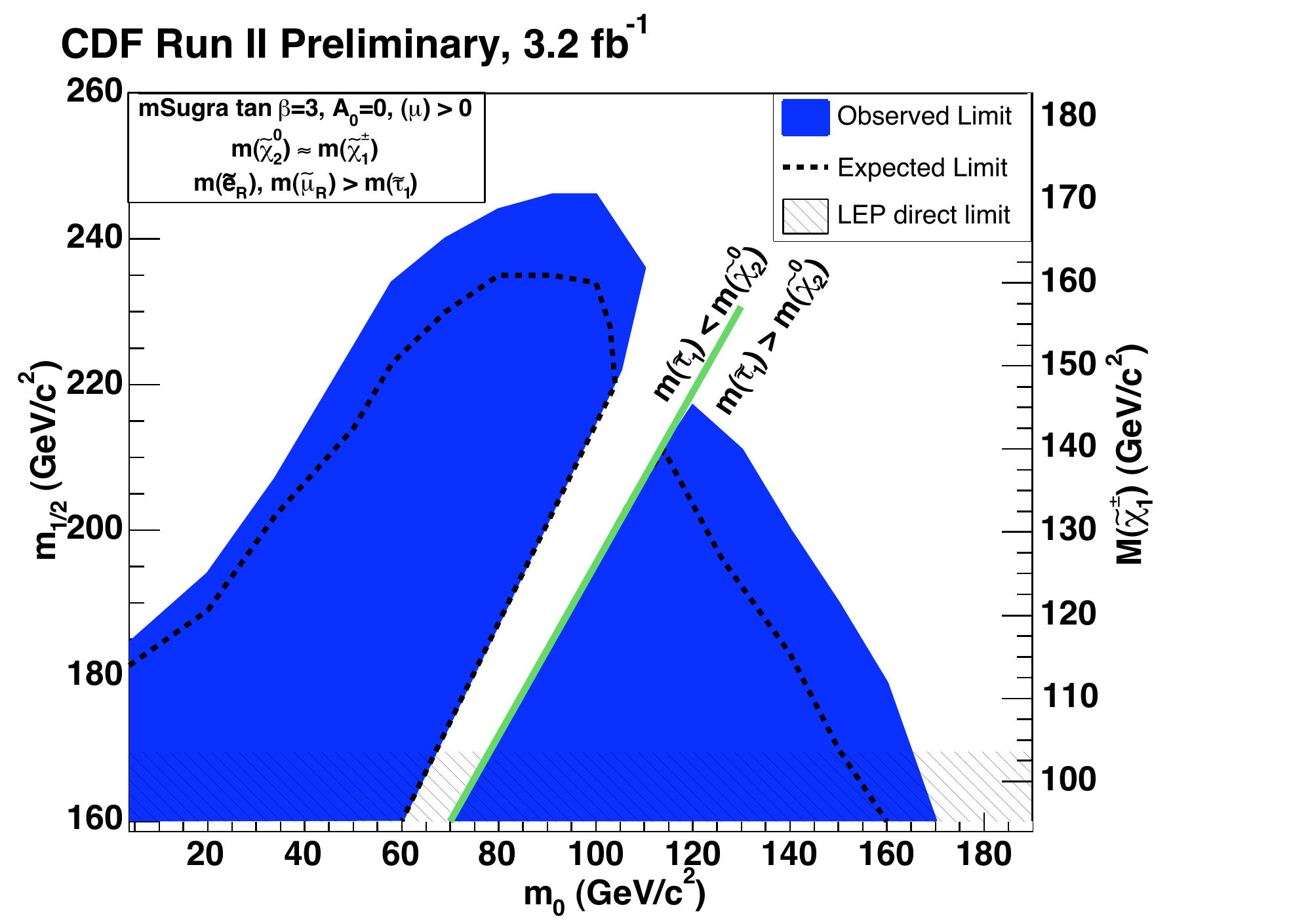}
\caption{One-dimensional chargino-neutralino cross-section vs. chargino-mass limit (left) and two-dimensional $m_{1/2}$ vs. $m_0$ limit (right) in the mSUGRA scenario.}
\end{center}
\end{minipage}  
\hfill
\end{figure}
Limits are calculated with a frequentist method \cite{trj}.  The chargino-mass 95\% confidence level
limit is 164 GeV/$c^2$ (155 GeV/$c^2$ expected) for a cross section of $\sim 0.1$ pb.
By varying both $m_0$ and $m_{1/2}$ we can set limits on these mSUGRA parameters,
as seen in Figure 2.

We are currently in the process
of improving these results by including very low-$p_T$ leptons ($> 5$ GeV/$c$ for all objects)
and low-$M_{\ell\ell}$ ($< 20$ GeV/$c^2$) -- continuing the work in \cite{strologas}) --
by including forward ($|\eta|>1$) leptons and also by including hadronically decaying tau leptons.  
These improvements not only increase
the event yield by threefold, but also allow us to investigate new regions in parameter and kinematic
phase space.  For example, the predominant decays of charginos and neutralinos through staus
(especially for high $\tan\beta$ values) will result in lower-$p_T$ leptons from the 
leptonic decays of the final-state tau leptons, and in hadronically decaying taus.  Cascade
decays of SUSY particles will also result in low-$p_T$ leptons and low dilepton masses.  This search
will maximize our discovery potential and sensitivity to SUSY parameter space.
\vspace{-0.2cm}

\bibliographystyle{aipproc}   

\end{document}